\documentstyle[preprint,prbbib,aps]{revtex}

\begin{document}
\title{Generalized canonical approach to gyrokinetic theory}
\author{P. Nicolini\thanks{%
Piero.Nicolini@cmfd.univ.trieste.it}$\ ^{1,2}\ $ and M. Tessarotto\thanks{%
M.Tessarotto@cmfd.univ.trieste.it}$\ ^{1,3}\ $}
\address{$\ ^{1}\ $Dipartimento di Scienze Matematiche,
Universit\`{a} degli Studi di Trieste (Italy).\\
$\ ^{2}\ $Istituto Nazionale di Fisica Nucleare, Sezione di Trieste (Italy).%
\\
$\ ^{3}\ $Consorzio di Magnetofluidodinamica, Universit\`{a} degli
Studi di Trieste (Italy). \\
}
\date{\today }
\maketitle

\begin{abstract}
We face the well-known gyrokinetic problem, which arises in the
description of the dynamics of a charged particle subject to fast
gyration for the presence of a strong electromagnetic field. The
customary approach to gyrokinetic theory, using canonical
variables or identifying them ``a posteriori'' by means of Darboux
theorem, leads to potential complications and ambiguities due to
the fact that canonical coordinates are field-related. Here we
propose an innovative formulation to construct gyrokinetic
canonical variables based on the introduction of a new definition
of canonical transformation. The new approach permits to shed
light on this often controversial issue.

\end{abstract}


\bigskip

\section{Introduction}

The ``gyrokinetic problem'' regards the description of the dynamics of a
charged particle in the presence of suitably ``intense'' electromagnetic
(EM) fields realized by means of appropriate perturbative expansions for the
its equations of motion. The expansions are usually performed with respect
to the ratio $\varepsilon =r_{L}/L<<1$, where $L$ and $r_{L}$ are
respectively a characteristic scale length of the EM fields and the
velocity-dependent particle Larmor radius $r_{L}=\frac{w}{\Omega _{s}},$
with $\Omega _{s\text{ }}=\frac{qB}{mc}$ the Larmor frequency and ${\bf w}$
the orthogonal component of a suitable particle velocity [see equation (\ref%
{last})]. The goal of gyrokinetic theory is to construct with prescribed
accuracy in $\varepsilon $ the so called ``gyrokinetic'' or ``guiding center
variables'', by means of a suitable ``gyrokinetic'' transformation, such
that the equations of motion result independent of the gyrophase $\phi $,
being $\phi $ the angle of fast gyration, which characterizes the motion of
a charged particle subject to the presence of a strong magnetic field.

The first author who systematically investigated the gyrokinetic problem was
probably Alfven\cite{Alfen 1950} who pointed out the existence of an
adiabatic invariant, the magnetic moment $\mu $, proportional to $p_{\phi }$
the conjugate canonical momentum to $\phi ,$ in the sense:
\begin{equation}
\frac{d}{dt}\ln \mu \sim O(\varepsilon ).
\end{equation}%
After subsequent work which dealt with direct construction methods of
gyrokinetic variables\cite{Gardner 1959,Gardner et al. 1959,Northrop et al.
1960,Lancina 1963,Lancina 1966,Krilin 1967a,Krilin 1967b,Dragt
1965,Bogolybov et al. 1961,Volsov 1962,Morozov 1966}, a significant step
forward was made by Kruskal\cite{Kruskal 1962} who, first, established the
consistency of the Alfven approach by proving, under suitable assumptions on
the EM fields, that the magnetic moment can be constructed correct at any
order $n$ in $\varepsilon $ in such a way that, denoting $M$ such a
dynamical variable, it results an adiabatic invariant of order $n,$ namely
in the sense%
\begin{eqnarray}
\frac{d}{dt}\ln M &\sim &O(\varepsilon ^{n}) \\
M &=&\mu +\varepsilon \mu _{1}+...+\varepsilon ^{n}\mu _{n}.
\end{eqnarray}%
A modern picture of the Hamiltonian formulation which makes easier the
formulation of higher order perturbative theories, was given only later by
Littlejohn\cite{Littlejohn1979} in terms of a non-canonical Lie-transform
approach, adopting a suitable set of noncanonical variables. As a motivation
to his noncanonical approach, Littlejohn\cite%
{Littlejohn1979,Littlejohn1981,Littlejohn1982,Littlejohn1983} pointed out
what in his views was a critical point of purely canonical formulations such
as previously developed Lie transform approaches\cite{Cary et al.
1977,Johnston et al. 1978}, namely the ambiguity in the separation of the
unperturbed and perturbed contributions in the Hamiltonian due the presence
of the vector potential ${\bf A}$ in the canonical momenta. He showed that
this difficulty can be circumvented by making use of suitable non-canonical
variables independent of ${\bf A}$ and which include the canonical pair $%
(\phi ,p_{\phi }).$

The possibility of constructing canonical gyrokinetic variables has relied,
since, on only two methods due respectively to Littlejohn\cite%
{Littlejohn1979,Littlejohn1981,Littlejohn1982,Littlejohn1983} and Gardner%
\cite{Gardner 1959,Gardner et al. 1959}.

The first approach, and probably the most popular in the literature\cite%
{White et al. 1984,Hahm Lee Brizard 1988,Hahm 1988,White 1990} is the based
on the use of Darboux theorem which allows, in principle, the construction
of canonical variables for an arbitrary differential 1-form. The canonical
1-form expressed in terms of the canonical variables is then obtained by
applying recursively the so-called ``Darboux reduction algorithm''as pointed
out by Littlejohn, which is obtained by a suitable combination of dynamical
gauge and coordinate transformations.

The second approach, due to Gardner\cite{Gardner 1959,Gardner et al. 1959},
was based on a mixed-variable generating function formulation. This was used
to construct the gyrokinetic canonical variables by means of a sequence of
canonical transformations. In particular, an attempt to extend this approach
to higher orders in $\varepsilon $ was later made by Weitzner \cite{Weitzner
1995}.

It should be stressed that in both approaches the canonical coordinates are
field-related, namely they depend on the particular geometry of the magnetic
field flux lines. As a consequence the construction of canonical variables
is achieved only under certain restrictions on the magnetic field (for the
Lie transform approach see in particular \ \cite{Hahm 1988,White 1990}). For
example, White\cite{White 1990} assumed only small deviations from the
axis-symmetric toroidal geometry, restricting himself to weakly chaotic
magnetic fields (assumption of regularity for the magnetic field), while
Weitzner\cite{Weitzner 1995}, based on his canonical approach, conjectured
that the explicit construction of the gyrokinetic canonical variables might
not be always possible. He claimed, indeed, that the magnetic moment might
not result single-valued for locally chaotic magnetic fields, such as those
occurring in non-symmetric MHD equilibria, for instance in Stellarators
(assumption of ``quasi-symmetric'' magnetic field \cite{Tessarotto 1995}).
An implication of his conjecture would be \ that for closed or ergodic
orbits the magnetic moment would not be\ any more an adiabatic invariant.
This conclusion, if proven correct, on one side might indicate a potential
breakdown of gyrokinetic theory itself and on the other might point out a
difficulty intrinsic to the field-related choice of the gyrokinetic
canonical coordinates. On the other hand, since no such assumption of
regularity is required for\ the magnetic field in customary noncanonical
formulations, this conjecture raises also a potential contradiction between
canonical and non-canonical formulations. An open issue is, therefore, the
possibility of constructing more general, field-geometry independent
canonical coordinates.

The goal of this paper is to propose a new solution to this problem. Our
approach is based on a new definition of canonical transformations, which we
call generalized canonical transformations. These transformations make use
of a suitable set of superabundant gyrokinetic variables, which include in
particular the canonical pair $\left( \phi ,p_{\phi }\right) $. Basic
feature of the new variables is that they can be defined independently of
the magnetic flux lines geometry and do not require the use of the Darboux
reduction algorithm.\ Basic consequences are that, on one hand, no such
regularity or quasi-symmetry assumptions are required for the magnetic
field, contrary to the conjecture of Weitzner, while, on the other hand, the
magnetic moment in these variables results in all cases a single-valued
function and therefore a physical observable.

The paper is organized as follows: we preliminary reformulate Hamiltonian
mechanics, extending it to the case of superabundant coordinates. Then, we
face the gyrokinetic problem, showing how a canonical transformation can be
written in terms of superabundant gyrokinetic coordinates, which are shown
to obey a suitable form of Hamilton modified variational principle.

\section{Generalized canonical transformations}

\bigskip

The purpose of this section is to extend the concept of canonical
transformation. In fact, the customary definition (see for example \cite%
{Goldstein 1980,Scheck 1994})\ may result, in some instances, too
restrictive for actual applications.

As is well known, the canonical transformation conventionally concerns
Hamiltonian systems $\left\{ {\bf x,\ }H({\bf x},t)\right\} $, natively
represented in terms of canonical variables ${\bf x=}({\bf q},{\bf p})$ and
obeying the Hamilton equations
\begin{equation}
\stackrel{\cdot }{{\bf x}}=\left[ {\bf x,}H({\bf x},t)\right]
\label{SymHamEqns}
\end{equation}%
being $H({\bf x},t)$ the a suitably regular Hamiltonian function. In the
phase space $\Gamma ,$ of dimension $2g$ (where $g$ is the degree of
freedom), spanned by the vector\ ${\bf x}=({\bf q},{\bf p}),$ a canonical
transformation is usually defined as a $C^{(n)}-$diffeomorphism (with $n\geq
2)$%
\begin{equation}
\gamma _{C}:{\bf x}\rightarrow \overline{{\bf x}}  \label{CanTrans}
\end{equation}%
to a phase space $\Gamma ^{\prime }$ having the same dimension of the
initial space $\Gamma $ and satisfying the symplectic condition
\begin{equation}
\underline{\underline{{\bf J}}}=\underline{\underline{{\bf M}}}\cdot
\underline{\underline{{\bf J}}}\cdot \underline{\underline{{\bf M}}}^{T}
\end{equation}%
where $\underline{\underline{{\bf M}}}=\frac{\partial \overline{{\bf x}}}{%
\partial {\bf x}}$ is the Jacobian matrix and $\underline{\underline{{\bf J}}%
}=\left(
\begin{array}{cc}
0 & I \\
-I & 0%
\end{array}%
\right) $ is the symplectic (Poisson) matrix of dimension $2g\times 2g.$

It is well-known that canonical transformations can be defined also in the
extended phase-space with dimension $2g+2,$ spanned by the vector $({\bf q,}%
q_{g+1}=t,{\bf p,}p_{g+1})$ [see \cite{Goldstein 1980}] and characterized by
a set of superabundant variables (canonical extended variables) which
satisfy Hamilton equations. Such extended transformations can be regarded as
a particular case of what we shall define as {\em generalized canonical
transformation}\label{trasformazione canonica generalizzata}, i.e., a $%
C^{(n)}$ diffeomorphism (with $n\geq 2$)%
\begin{equation}
\gamma _{G}:{\bf x}\rightarrow {\bf x}_{G}={\bf x}_{G}({\bf x},t,\alpha ),
\label{GenCanTrans}
\end{equation}%
(with $\alpha $ a real parameter) for which the transformed phase-space $%
\Gamma _{G}$ can have larger dimension of the initial phase space $\Gamma $
and is characterized by superabundant variables, subject either to Hamilton
or finite terms constraint equations. More precisely, first we assume $\dim
(\Gamma _{G})=2g^{\prime }+k>\dim (\Gamma )=2g,$ with $g^{\prime }\geq g$
and letting ${\bf x}_{G}$ of the form ${\bf x}_{G}=({\bf z},{\bf u}),$ with $%
{\bf z}=\left( z_{1},...z_{2g^{\prime }}\right) $ and ${\bf u}=\left(
u_{1},...u_{k}\right) .$ Then, we require that ${\bf z}$ and ${\bf u}$ obey
respectively the extended Hamilton equations (for $i=1,...,2g^{\prime })$%
\begin{equation}
\frac{d}{dt}z_{i}({\bf x},t,\alpha )=\sum_{j=1,2g^{\prime }}J_{ij}^{\prime
}\cdot \frac{\partial }{\partial z_{j}}K({\bf x}_{G},t,\alpha ),
\label{GenSymEqns}
\end{equation}%
(with $J_{ij}^{\prime }$ the canonical Poisson tensor of rank $2g^{\prime }$
and assuming that the temporal variable $t$ is left invariant by the
transformation$)$ and $k$ constraint equations of the form
\begin{equation}
f_{s}({\bf z},{\bf u},t)=0,  \label{GenCons}
\end{equation}%
with $s=1,k$ and $f_{s}({\bf z},{\bf u},t)$ real $C^{(2)}$ functions.

The components of the transformed state ${\bf x}_{G}$ will be here denoted
as {\em superabundant canonical variables}%
\index{variabili canoniche!variabili canoniche essenziali} and the
corresponding ``equations of motion'' (\ref{GenSymEqns}), (\ref{GenCons})
generalized canonical equations.

It is immediate to point out that generalized canonical equation can be set
in variational form in terms of a suitable constrained form of modified
Hamilton variational principle, just as the usual canonical equations. To
provide a straightforward example, which will be also used in the sequel,
let us consider the case of a charged point particle subject to an EM field
and defined the following generalized canonical transformation
\begin{equation}
{\bf x=(r,p)}\rightarrow {\bf x}_{G}=({\bf r,p,v}).
\end{equation}%
One can prove that that the superabundant state ${\bf x}_{G}$ is an extremal
curve of the action functional
\begin{equation}
S({\bf x}_{G})=\int_{t_{1}}^{t_{2}}dt%
\widehat{{\cal L}}({\bf x}_{G},\stackrel{\cdot }{{\bf r}},t)
\end{equation}%
with\ the fundamental 1-form:%
\begin{equation}
\widehat{{\cal L}}({\bf x}_{G},\stackrel{\cdot }{{\bf r}}{\bf ,}t)dt=d{\bf r}%
\cdot {\bf p}\ -dtH({\bf r,p},t)-\left[ d{\bf r}-{\bf v}dt\right] \cdot %
\left[ {\bf p}\ -m{\bf v-}\frac{q}{c}{\bf A}\right] .  \label{Lhat}
\end{equation}%
where the Hamiltonian is%
\begin{equation}
H({\bf r,p},t)=\frac{1}{2m}\left[ {\bf p}-\frac{q}{c}{\bf A}\right]
^{2}+q\Phi
\end{equation}%
(with $\left\{ {\bf A},\Phi \right\} $ the EM potentials). It follows that
the corresponding Euler-Lagrange (E-L) equations coincide with the canonical
equations%
\begin{equation}
\stackrel{\cdot }{{\bf x}}=\left[ {\bf x},H\right]
\end{equation}%
plus the non-holonomic constraint%
\begin{equation}
{\bf p}=m{\bf v}-\frac{q}{c}{\bf A.}
\end{equation}%
Therefore ${\bf x}_{G}$ defines a generalized canonical state for the
Hamiltonian system.

\section{Hamiltonian gyrokinetic theory in superabundant variables}

In this section we intend to show\ that the previous Hamiltonian system,
under a suitable assumption of ``strong'' EM field, can be represented by
means of a appropriate set of superabundant canonical variables which are
gyrokinetic, i.e. the new Hamiltonian results independent of the gyrophase $%
\phi $ when expressed in terms of them. The other basic feature is that the
the corresponding canonical coordinates can always be chosen to be
field-independent. To construct the new gyrokinetic variables we shall
follow a two-step approach.

The first step consists in constructing a particular set of hybrid\footnote{%
In some cases it is convenient to adopt, for Hamiltonian or Lagrangian
systems, a set of variables, which are not Hamiltonian or Lagrangian or
Newtonian. \ In such circumstances we shall speak of hybrid variables.}
superabundant gyrokinetic variables, here denoted as {\it pseudo-canonical} $%
{\bf x}^{\prime }\equiv ({\bf r}^{\prime },p_{{\bf r}^{\prime }},\phi
^{\prime },p_{\phi ^{\prime }}).$ For definiteness let us require that the
Hamiltonian function takes the form%
\begin{equation}
H({\bf r,p},t)=\frac{1}{2m}\left[ {\bf p}-\frac{q}{\varepsilon c}{\bf A}%
\right] ^{2}+\frac{q}{\varepsilon }\Phi
\end{equation}%
where $\varepsilon $ is a real infinitesimal and assume that the EM
potentials $\Phi ,\ {\bf A}$ \ are analytic functions of $\varepsilon $ and
can be represented in the form%
\begin{eqnarray}
\Phi  &=&\sum_{i=-1}^{\infty }\varepsilon ^{i}\Phi _{i}({\bf r},t) \\
{\bf A} &=&\sum_{i=-1}^{\infty }\varepsilon ^{i}{\bf A}_{i}({\bf r},t).
\end{eqnarray}%
In validity of this assumption the construction of hybrid gyrokinetic
variables is well known and has been achieved by several authors (see for
example\cite{Littlejohn1979,Littlejohn1981,Littlejohn1982,Littlejohn1983}).
In this case the Lagrangian expressed in terms of gyrokinetic variables
(gyrokinetic Lagrangian) reads
\begin{equation}
\overline{{\cal L}}({\bf y}^{\prime },\stackrel{\cdot }{{\bf r}}^{\prime },%
\stackrel{\cdot }{\phi }^{\prime },t)=\stackrel{\cdot }{{\bf r}}^{\prime }%
{\bf \cdot }\frac{q}{\varepsilon c}{\bf A}^{\ast }({\bf r}^{\prime
},u^{\prime },\mu ^{\prime },t)+
\end{equation}%
\begin{equation}
-\left( \frac{\stackrel{.}{\phi }^{\prime }}{\Omega _{s}^{\prime }}+1\right)
\mu ^{\prime }B^{\prime }-\frac{m}{2}{\bf v}^{\prime 2}-\frac{q}{\varepsilon
}\Phi ^{\ast }({\bf r}^{\prime },u^{\prime },\mu ^{\prime },t),
\label{Lagrangiana gyro}
\end{equation}%
where the hybrid state ${\bf y}^{\prime }$ is defined as follows
\begin{equation}
{\bf y}^{\prime }\equiv \left( {\bf r}^{\prime },u^{\prime },\mu ^{\prime
},\phi ^{\prime }\right) ,
\end{equation}%
and, ignoring for sake of simplicity\footnote{%
In the sequel we shall omit higher order correction in $\varepsilon .$}
corrections of order $O(\varepsilon ^{2})$, one can directly prove that
there results as a consequence
\begin{equation}
{\bf v}^{\prime }({\bf r}^{\prime },u^{\prime },\mu ^{\prime },t)\equiv
u^{\prime }{\bf b}^{\prime }+{\bf v}_{E}^{\prime }+\varepsilon {\bf v}%
_{D}^{\prime },
\end{equation}%
where ${\bf v}_{E}^{\prime }=c{\bf E}^{\prime }\times {\bf b}^{\prime
}/B^{\prime }$ is the electric drift velocity and ${\bf v}_{D}^{\prime }=%
\frac{{\bf b}^{\prime }}{\Omega ^{\prime }}\times \left\{ \frac{\mu ^{\prime
}}{m}\nabla ^{\prime }B^{\prime }+(u^{\prime }{\bf b}^{\prime }+{\bf v}%
_{E}^{\prime })\cdot (u^{\prime }\nabla ^{\prime }{\bf b}^{\prime }+\nabla
{\bf v}_{E}^{\prime })\right\} $ is the diamagnetic drift velocity, both
evaluated at the guiding center position. Here the notations are standard.
Thus ${\bf b}^{\prime }={\bf B}({\bf r}^{\prime },t)/B({\bf r}^{\prime },t)$
while the primes denote quantities evaluated at the guiding center position $%
{\bf r}^{\prime }.$ In particular, $\mu ^{\prime }$ is the magnetic moment
evaluated at the guiding center position. Moreover, $\left\{ {\bf A}^{\ast
},\Phi ^{\ast }\right\} $ are the effective EM potentials, which at first
order in $\varepsilon $ read
\begin{eqnarray}
{\bf A}^{\ast }({\bf r}^{\prime },u^{\prime },w^{\prime },t) &=&{\bf A}%
^{\prime }+\frac{\varepsilon mc}{q}{\bf v}^{\prime }\left[ 1+O(\varepsilon )%
\right] ,  \label{potenziale vettore efficace} \\
\Phi ^{\ast }({\bf r}^{\prime },u^{\prime },w^{\prime },t) &=&\Phi ^{\prime }%
\left[ 1+O(\varepsilon )\right] .
\end{eqnarray}%
\index{formulazione pseudo-Hamiltoniana!formulazione girocinetica} To
construct a set of superabundant variables, let us introduce the conjugate
momenta
\begin{equation}
p_{{\bf r}^{\prime }}\ =%
\frac{\partial \overline{{\cal L}}}{\partial \left( \frac{d}{dt}{\bf r}%
^{\prime }\right) }=\frac{q}{\varepsilon c}{\bf A}^{\ast }\equiv m{\bf v}%
^{\prime }{\bf +}\frac{q}{\varepsilon c}{\bf A}^{\prime },
\label{momento canonico girocinetico}
\end{equation}%
\begin{equation}
p_{\phi ^{\prime }}=\frac{\partial \overline{{\cal L}}}{\partial \left(
\frac{d}{dt}\phi ^{\prime }\right) }=-\frac{1}{\Omega _{s}^{\prime }}\mu
^{\prime }B^{\prime }=-\frac{mc}{q}\mu ^{\prime },
\label{secondo momento can. girocinetico}
\end{equation}%
in terms of which the gyrokinetic Lagrangian becomes%
\begin{equation}
\overline{{\cal L}}({\bf x}^{\prime },\stackrel{\cdot }{{\bf r}}^{\prime },%
\stackrel{\cdot }{\phi }^{\prime },t)=\stackrel{\cdot }{{\bf r}}^{\prime }%
{\bf \cdot }p_{{\bf r}^{\prime }}+\stackrel{\cdot }{\phi }^{\prime }p_{\phi
^{\prime }}-K({\bf x}^{\prime },t).
\end{equation}%
The superabundant state ${\bf x}^{\prime }\equiv ({\bf r}^{\prime },p_{{\bf r%
}^{\prime }},\phi ^{\prime },p_{\phi ^{\prime }})$ is here denoted as
pseudo-canonical and $K({\bf x}^{\prime },t)$ is the corresponding
Hamiltonian
\begin{equation}
K({\bf x}^{\prime },t)=-p_{\phi ^{\prime }}\Omega +\frac{1}{2m}\left[ p_{%
{\bf r}^{\prime }}\ -\frac{q}{\varepsilon c}{\bf A}^{\prime }\right] ^{2}+%
\frac{q}{\varepsilon }\Phi ^{\ast }.
\end{equation}%
Manifestly, the transformation
\begin{equation}
{\bf x=}\left( {\bf r,}p_{{\bf r}}\right) \rightarrow {\bf x}^{\prime }{\bf =%
}\left( {\bf r}^{\prime }{\bf ,}p_{{\bf r}^{\prime }},\phi ^{\prime
},p_{\phi ^{\prime }}\right)   \label{trasf.0}
\end{equation}%
is not canonical, even in the generalized sense previously indicated.

The second step concerns the introduction of a further transformation to a
new set of superabundant gyrokinetic variables. In particular let us
consider the transformation
\begin{equation}
\gamma :{\bf x=}\left( {\bf r,}p_{{\bf r}}\right) \rightarrow {\bf X}%
^{\prime }{\bf =}\left( {\bf r}^{\prime }{\bf ,}p_{{\bf r}^{\prime }},\phi
^{\prime },p_{\phi ^{\prime }},{\bf v}^{\prime }\right)
\end{equation}%
where ${\bf v}^{\prime }\equiv u^{\prime }{\bf b}^{\prime }+{\bf v}%
_{E}^{\prime }+\varepsilon {\bf v}_{D}^{\prime }$ is here considered as an
independent variable. We intend to prove that the gyrokinetic state ${\bf X}%
^{\prime }$ is canonical in the generalized sense defined above, namely its
components satisfy either Hamilton equations with respect to a suitable
Hamiltonian function or finite-terms constraint conditions. To reach the
proof, we initially notice that, by construction [see. Eq.(\ref{momento
canonico girocinetico})], the following finite-terms constraint is satisfied
by the vector ${\bf v}^{\prime }$
\begin{equation}
\frac{1}{m}\left[ p_{{\bf r}^{\prime }}\ -\frac{q}{\varepsilon c}{\bf A}%
^{\prime }\right] ={\bf v}^{\prime }.  \label{vincolo giurocinetico 0}
\end{equation}%
Furthermore, using $\left( {\bf r}^{\prime },u^{\prime },p_{\phi ^{\prime
}},\phi ^{\prime }\right) $ the E-L equations for ${\bf r}^{\prime }$ reads,
omitting higher order terms in $\varepsilon $
\begin{equation}
-\frac{d}{dt}p_{{\bf r}^{\prime }}+m\left( u^{\prime }\nabla ^{\prime }{\bf b%
}^{\prime }+\nabla ^{\prime }{\bf v}_{E}^{\prime }+\frac{q}{\varepsilon mc}%
\nabla ^{\prime }{\bf A}^{\prime }\right) \cdot \stackrel{\cdot }{{\bf r}}%
^{\prime }+  \label{eq.per pr}
\end{equation}%
\begin{equation}
+p_{\phi ^{\prime }}\nabla ^{\prime }\Omega -m\left( u^{\prime }\nabla
^{\prime }{\bf b}^{\prime }+\nabla ^{\prime }{\bf v}_{E}^{\prime }\right)
\cdot \left[ u^{\prime }{\bf b}^{\prime }+{\bf v}_{E}^{\prime }\right] -%
\frac{q}{\varepsilon }\nabla ^{\prime }\Phi ^{\ast }={\bf 0},
\end{equation}%
from one can prove that it follows%
\begin{equation}
\stackrel{\cdot }{{\bf r}}^{\prime }={\bf v}^{\prime }.
\label{vincolo girocinetico}
\end{equation}%
As a consequence, there results
\begin{equation}
-\frac{d}{dt}p_{{\bf r}^{\prime }}+\frac{q}{\varepsilon c}\nabla ^{\prime }%
{\bf A}^{\prime }\cdot {\bf v}^{\prime }+p_{\phi ^{\prime }}\nabla ^{\prime
}\Omega -\frac{q}{\varepsilon }\nabla ^{\prime }\Phi ^{\ast }=0,
\end{equation}%
which can be cast in Hamiltonian form with respect to $K({\bf x}^{\prime },t)
$
\begin{equation}
\frac{d}{dt}p_{{\bf r}^{\prime }}=-\frac{\partial }{\partial {\bf r}^{\prime
}}K({\bf x}^{\prime },t),  \label{prHeqns}
\end{equation}%
where the partial derivative $\partial /\partial {\bf r}^{\prime }$ is
defined keeping $p_{{\bf r}^{\prime }}$ as a constant.\ Analogously, from (%
\ref{vincolo giurocinetico 0}) and (\ref{vincolo girocinetico}) one obtains
the Hamilton equations
\begin{equation}
\frac{d}{dt}{\bf r}^{\prime }=\frac{\partial }{\partial p_{{\bf r}^{\prime }}%
}K({\bf x}^{\prime },t)=\frac{1}{m}\left[ p_{{\bf r}^{\prime }}\ -\frac{q}{%
\varepsilon c}{\bf A}^{\prime }\right] .  \label{rHeqns}
\end{equation}%
Finally, in a similar way, it is immediate to prove that also $p_{\phi
}^{\prime }$ and $\phi ^{\prime }$ obey Hamilton equations
\begin{equation}
\frac{d}{dt}p_{\phi }^{\prime }=-\frac{\partial }{\partial \phi ^{\prime }}K(%
{\bf x}^{\prime },t)=0,  \label{pphiHequns}
\end{equation}%
\begin{equation}
\frac{d}{dt}\phi ^{\prime }=\frac{\partial }{\partial p_{\phi }^{\prime }}K(%
{\bf x}^{\prime },t)=-\Omega _{s}^{\prime }.  \label{phiHeqns}
\end{equation}%
Thus, given the constraint condition (\ref{vincolo giurocinetico 0}), which
establishes a finite-terms equation for ${\bf v}^{\prime },$ the remaining
variables ${\bf x}^{\prime }{\bf =}\left( {\bf r}^{\prime }{\bf ,}p_{{\bf r}%
^{\prime }},\phi ^{\prime },p_{\phi ^{\prime }}\right) $ satisfy the
Hamilton equations, with respect to the Hamiltonian function $K({\bf x}%
^{\prime },t)$. As a consequence,\ the superabundant state ${\bf X}^{\prime }%
{\bf =}\left( {\bf r}^{\prime }{\bf ,}p_{{\bf r}^{\prime }},\phi ^{\prime
},p_{\phi ^{\prime }},{\bf v}^{\prime }\right) $ establishes a generalized
canonical transformation [see for example (\ref{GenCanTrans})] in the
transformed phase-space $\Gamma _{G}.$ Furthermore\ ${\bf X}^{\prime }$
results, by construction, a gyrokinetic state and therefore the gyrophase $%
\phi $ is ignorable for the generalized canonical equations (\ref{vincolo
giurocinetico 0}, \ref{prHeqns}, \ref{rHeqns}, \ref{pphiHequns}, \ref%
{phiHeqns}).

The crucial feature of these variables is that the canonical coordinates $%
{\bf r}^{\prime }$ are manifestly independent of any particular magnetic
field geometry, which implies that the vector ${\bf r}^{\prime }$ can be
represented in the form ${\bf r}^{\prime }={\bf r}^{\prime }({\bf q}^{\prime
},t)$,\ being ${\bf q}^{\prime }=(q_{1}^{\prime },q_{2}^{\prime
},q_{3}^{\prime })$\ arbitrary, field-independent, curvilinear coordinates,
such as for instance orthogonal Cartesian coordinates. As a consequence, no
restriction is placed on the magnetic field geometry for the definition of
these canonical variables, contrary to previous formulations\cite{White et
al. 1984,White 1990}. A fundamental consequence is that, by construction,
the magnetic moment $\mu $ can always be defined in such a way to be a
single-valued function with respect with to any angle-like coordinates ${\bf %
q}^{\prime }=(q_{1}^{\prime },q_{2}^{\prime },q_{3}^{\prime })$\ and
therefore results, in a suitable gauge, a physical observable. In
particular, contrary to the conjecture of Weitzner \cite{Weitzner 1995}, the
definition of the magnetic moment results independent of the magnetic field
topology and does not require the existence of a single family of nested
magnetic surfaces (quasi-symmetric magnetic field) \cite{Tessarotto 1995}.

\section{Constrained Hamilton modified variational principle}

A further key aspect of the present approach is that the previous
generalized canonical equations (namely \ref{vincolo giurocinetico 0}, \ref%
{prHeqns}, \ref{rHeqns}, \ref{pphiHequns}, \ref{phiHeqns}) are necessarily
variational, as pointed out in section II. In fact, it is easy to show that
they follow from a constrained form of Hamilton modified variational
principle. One can prove that this is provided in terms of the following
gyrokinetic Lagrangian:
\begin{equation}
{\cal L}^{\prime }({\bf X}^{\prime }{\bf ,}\stackrel{\cdot }{{\bf r}}%
^{\prime },\stackrel{.}{\phi }^{\prime },t)=\stackrel{\cdot }{{\bf r}}%
^{\prime }\cdot p_{{\bf r}^{\prime }}\ +\stackrel{.}{\phi }^{\prime }p_{\phi
^{\prime }}\ -\overline{{\cal K}}({\bf x}^{\prime },t)-
\label{L girocinetica}
\end{equation}%
\[
-\left[ \stackrel{\cdot }{{\bf r}}^{\prime }-{\bf v}^{\prime }\right] \cdot %
\left[ p_{{\bf r}^{\prime }}\ -m{\bf v}^{\prime }{\bf -}\frac{q}{%
c\varepsilon }{\bf A}^{\prime }\right] ,
\]%
\bigskip where $\overline{{\cal K}}({\bf x}^{\prime },t)$ is the
corresponding Hamiltonian
\begin{equation}
\overline{{\cal K}}({\bf x}^{\prime },t)=\frac{1}{2m}\left[ p_{{\bf r}%
^{\prime }}\ -\frac{q}{\varepsilon c}{\bf A}^{\prime }\right] ^{2}-\frac{1}{%
\varepsilon }\Omega _{s}^{\prime }p_{\phi ^{\prime }}+\frac{q}{\varepsilon }%
\Phi ^{\prime }
\end{equation}%
and by definition ${\bf v}^{\prime }\equiv u^{\prime }{\bf b}^{\prime }+{\bf %
v}_{E}^{\prime }+\varepsilon {\bf v}_{D}^{\prime }.$

The E-L equations corresponding to (\ref{L girocinetica}) coincide with the
previous equations of motion (\ref{vincolo giurocinetico 0}, \ref{prHeqns}, %
\ref{rHeqns}, \ref{pphiHequns}, \ref{phiHeqns}) and hence provide a
variational formulation for them.

Finally, we emphasize that the constrained Lagrangian (\ref{L girocinetica})
defined above can also be obtained directly from the Lagrangian of a charged
point particle in the presence of a strong EM field previously defined [see
equation (\ref{Lhat})]. This is achieved by means of the gyrokinetic
transformation defined in terms of generalized canonical variables, i.e. of
the form%
\begin{eqnarray}
{\bf r} &=&{\bf r}^{\prime }+\varepsilon {\bf r}_{1}+... \\
p_{{\bf r}} &=&p_{{\bf r}^{\prime }}^{\prime }+\varepsilon p_{{\bf r}%
_{1}}+... \\
{\bf v} &=&{\bf v}^{\prime }+\varepsilon {\bf v}_{1}+...
\end{eqnarray}%
with perturbations $\varepsilon {\bf r}_{1},\varepsilon p_{{\bf r}%
_{1}},\varepsilon {\bf v}_{1}$ to be suitably defined. This transformation
can be constructed using standard techniques (see for example \cite%
{Littlejohn1979,Littlejohn1981,Littlejohn1982,Littlejohn1983}) and can be
identified at any order in $\varepsilon $ with a generalized canonical
transformation. In particular, to the leading order in $\varepsilon $ one
can prove

\begin{equation}
\left\{
\begin{array}{c}
{\bf r} \\
p_{{\bf r}} \\
{\bf v}%
\end{array}%
\right. \rightarrow \left\{
\begin{array}{c}
{\bf r}={\bf r}^{\prime }+\varepsilon {\bf \rho }^{\prime }, \\
p_{{\bf r}}=p_{{\bf r}^{\prime }}\ +m{\bf w}^{\prime }+\frac{q}{c}{\bf \rho }%
^{\prime }{\bf \cdot \nabla }^{\prime }{\bf A}^{\prime }, \\
{\bf v=v}^{\prime }+{\bf w}^{\prime }{\bf ,}%
\end{array}%
\right.
\end{equation}%
where $\varepsilon {\bf \rho }$ denotes the Larmor radius
\begin{equation}
\varepsilon {\bf \rho =-\varepsilon }\frac{{\bf w}^{\prime }{\bf \times b}%
^{\prime }}{\Omega ^{\prime }}.
\end{equation}%
The remaining notation is standard. Thus we require
\begin{equation}
\left\{
\begin{array}{c}
{\bf v}^{\prime }=u^{\prime }{\bf b}^{\prime }+{\bf v}_{E}^{\prime
}+\varepsilon {\bf v}_{D}^{\prime }, \\
{\bf w}^{\prime }{\bf =}w^{\prime }\left( {\bf e}_{1}^{\prime }\cos \phi
^{\prime }+{\bf e}_{2}^{\prime }\sin \phi ^{\prime }\right) , \\
\phi ^{\prime }=arctg\left\{ \frac{\left( {\bf v-v}_{E}^{\prime }\right)
\cdot {\bf e}_{2}^{\prime }}{\left( {\bf v-v}_{E}^{\prime }\right) \cdot
{\bf e}_{1}^{\prime }}\right\} , \\
w^{\prime }\equiv \sqrt{2B^{\prime }\mu ^{\prime }}, \\
p_{\phi ^{\prime }}\ =-\frac{mc}{q}\mu ^{\prime },%
\end{array}%
\right.  \label{last}
\end{equation}%
where ${\bf w}^{\prime }$ is a vector in the plane orthogonal to the
magnetic flux line. Therefore, by means of the generalized canonical
variables ${\bf X}^{\prime }{\bf =}\left( {\bf r}^{\prime }{\bf ,}p_{{\bf r}%
^{\prime }},\phi ^{\prime },p_{\phi ^{\prime }},{\bf v}^{\prime }\right) $
gyrokinetic theory can be directly represented in canonical form without
recurring to the use of hybrid gyrokinetic variables as considered by most
of previous authors \cite%
{Littlejohn1979,Littlejohn1981,Littlejohn1982,Littlejohn1983,Cary et al.
1977,Johnston et al. 1978,White et al. 1984,Hahm Lee Brizard 1988,Hahm
1988,White 1990,Weitzner 1995}

\section{Final remarks\protect\bigskip\ }

Based on the concept of generalized canonical transformation a new set of
superabundant canonical variables has been defined for the set of
gyrokinetic variables, which can be used to represent the Lagrangian of a
charged point particle immersed in a strong EM field. These canonical
variables, contrary to the customary ones previously considered in the
literature \cite{Gardner 1959,Gardner et al. 1959,White 1990,Weitzner 1995},
do not depend on the magnetic field geometry and do not require subsidiary
restrictions for the magnetic field. As a consequence, gyrokinetic theory
can be developed in principle at any order in $\varepsilon ,$ by means of
gyrokinetic transformation expressed in terms of suitable generalized
canonical variables. The present formalism for its straightforward
simplicity and handy construction appears susceptible of interesting new
applications both in plasma theory and mathematical physics.

\subsection*{Acknowledgement}

P.N. is supported by MIUR PRIN project ``Metodi matematici nelle teorie
cinetiche'' and partially by INDAM (Istituto Nazionale di Alta Matematica),
through GNFM (Gruppo Nazionale di Fisica Matematica) and by INFN (Istituto
Nazionale di Fisica Nucleare), Sezione di Trieste (Italy). M.T. is also
partially supported by CMFD Consortium (Consorzio di Magnetofludodinamica),
Trieste (Italy).

\end{document}